\documentstyle[12pt]{article}

\thicklines                         % thick straight lines for pictures (LaTeX)
\def\a{\alpha}

\def\d{\delta}
\def\e{\epsilon}                % Also, \varepsilon
\def\f{\phi}                    %       \varphi
\def\g{\gamma}
\def\h{\eta}

\def\j{\psi}

\def\m{\mu}
\def\n{\nu}

\def\p{\pi}                     % Also, \varpi
\def\th{\theta}                  %       \vartheta
\def\s{\sigma}                  %       \varsigma

\def\D{\Delta}

\def\L{\Lambda}

\def\P{\Pi}

\def\ct{{\cal T}}

\def\cbo{{\,\raise-.15ex\Sc [\,}}                       % curly "
\def\Tilde#1{\widetilde{#1}}                    % big tilde
\def\Bar#1{\overline{#1}}                       % big bar
\def\svev#1{\left\langle #1\right\rangle}       % variable < >
\def\ddt#1{{\buildrel {\hbox{\LARGE .\kern-2pt.}} \over {#1}}}% double dot-over
\def\beqn#1{ \renewcommand{\theequation}{#1}
             \begin{eqnarray} }
\def\eeqn{ \renewcommand{\theequation}{\arabic{equation}}
           \end{eqnarray} }

\def\beqr#1{ \setcounter{equation}{#1}
             \begin{eqnarray} }

\def\eeqr{\end{eqnarray}}
\def\NON{\nonumber\\}
\def\beqrabc#1{ \setcounter{equation}{0}
                \renewcommand{\theequation}{#1\alph{equation}}
                \begin{eqnarray} }
\def\beqrn#1#2{ \setcounter{equation}{#2}
                \renewcommand{\theequation}{#1.\arabic{equation}}
                \begin{eqnarray} }

\def\NPB#1{Nucl. Phys. {\bf B#1}}
\def\NPBP#1{Nucl. Phys. (Proc. Suppl.) {\bf B#1}}
\def\PLB#1{Phys. Lett. {\bf B#1}}
\def\PRD#1{Phys. Rev. {\bf D#1}}

\def\PRL#1{Phys. Rev. Lett. {\bf #1}}

\def\sstyle{\scriptstyle}

\def\rhs{\mbox{r.h.s.} }

\def\ie{\mbox{i.e.} }

\def\leqx{\,\raisebox{-1.0ex}{$\stackrel{\textstyle <}{\sim}$}\,}

\def\frac#1#2{ {\sstyle {#1\over #2} } }

\def\tr{{\rm tr}\,}

\def\half{{1\over 2}}

\begin{document}
\noindent  \hfill WIS--93/99--OCT--PH
\par
\begin{center}
\vspace{15mm}
{\large\bf
Anomalies and chiral defect fermions}\\[5mm]
{\it by}\\[5mm]
Yigal Shamir\\
Department of Physics\\
Weizmann Institute of Science, Rehovot 76100, ISRAEL\\[15mm]
{ABSTRACT}\\[2mm]
  \end{center}
\begin{quotation}
  Chiral defect fermions in the background of an external, $2n$
dimensional gauge field are considered. Assuming first a finite extra
dimension, we calculate the axial anomaly
in a vector-like, gauge invariant model for arbitrary $n$,
and the consistent anomaly in a gauge {\it variant} model with a chiral
spectrum. For technical reasons,
the latter calculation is limited to the $2+1$
dimensional case. We also show that the infinite lattice chiral model,
when properly defined, is in fact a limiting case of the above gauge-variant
model. The behaviour of this model with a dynamical gauge field is
discussed.
\end{quotation}

\newpage
\noindent {\bf 1.~~ Introduction}
\vspace{3ex}

  Recently, interest in the long standing problem of constructing
chiral gauge theories on the lattice was revived~[1-12] following the
interesting proposal of D.~Kaplan~[1]. Kaplan starts with a Wilson
action in $2n+1$ dimensions, with the peculiar property that the Dirac
mass has the shape of a domain wall, \ie the mass term
changes sign across the $2n$ dimensional surface $s=0$. For a certain
range of the Dirac mass and the Wilson parameter, the spectrum
contains a single, $2n$ dimensional chiral fermion that lives on the
domain wall.

  Kaplan suggested that a chiral gauge theory in $2n$ dimensions
may emerge in the continuum limit of his model when dynamical
gauge fields are introduced.  His suggestion was based on the
expectation that it may be possible to decouple all but the
massless chiral mode in the low energy limit.

  There are several ways in which one can proceed from the free
domain wall model to a full fledged interacting theory.
In particular, the gauge field may be chosen to
be $2n+1$ dimensional, as originally proposed by Kaplan, or it can be
made $2n$ dimensional by eliminating its extra component and insisting
that the remaining $2n$ components be independent of the extra
coordinate~[3,4]. In this case the extra coordinate becomes a
sophisticated flavour space. We will restrict our attention in this
paper to the latter possibility. For recent results on
models with a $2n+1$ dimensional gauge field see ref.~[5].

  A basic difficulty  is how to
define the {\it interacting} model with an infinite extra direction as the
limit of models with a finite extra direction. The origin of this
difficulty is
as follows.  If the extra coordinate $s$ has an infinite range, the chiral
fermion on the domain wall is the only massless excitation.
On the other hand, if the extra coordinate has a finite range there
appears another massless mode with the opposite chirality.
For example, choosing periodic boundary conditions
necessitates the existence of an anti-domain wall, and the chiral
fermion on the anti-domain wall has the opposite chirality.
More examples of this phenomenon are discussed below.

  The successful construction of a chiral gauge theory using Kaplan's
method therefore depends critically on one's ability to decouple the
extra, unwanted massless mode. Especially stringent constraints exist
if  one attempts to achieve this decoupling at the level of perturbation
theory, because here the
Nielsen-Ninomiya theorem~[13] is directly  applicable.
Once the extra coordinate is made finite, and as long as gauge invariance
is maintained, both massless modes couple with equal strength to the
gauge field and the theory is vector-like. A sequence of such models
with increasing size of the extra direction will eventually give rise
to a vector-like theroy as well.

  In an attempt to avoid this problem from the outset, Narayanan and
Neuberger have proposed to work directly with an infinite extra
direction~[4].
The spectrum then contains a single chiral fermion, and at least
within the context for perturbation theory, one can investigate
the model by writing
down the corresponding Feynman rules. However, a closer look reveals
that the Feynman rules of the infinite lattice model are ambiguous.
Different values for the divergence of the source current are obtained
if the infinite summations over $s$ at the vertices of the relevant
diagram are carried out in different ways.

  Narayanan and Neuberger circumvented this difficulty by showing that
a chiral effective action can be defined using  transfer matrix
methods~[6]. In this approach, the
transfer matrix for a single site translation in the $s$-direction is
gauge invariant, and possible breakdown of gauge invariance resides
entirely in the choice of the boundary conditions.
However, the relation between the transfer matrix construction and the
original, infinite lattice model has not been fully clarified.

  In this paper we reexamine the infinite lattice chiral model directly.
The crucial property of a consistent set of Feynman rules is that,
during the process of
sending the range of the $s$-summations to infinity, all vertices in a given
diagram should describe the coupling of the gauge field to the {\it same}
source current. (In other words, the IR
regularized diagram should satisfy Bose symmetry). Enforcing this
requirement uniquely determines how to
perform the infinite $s$-summations. Moreover, it implies that the infinite
lattice chiral model is defined as the limit of a sequence of models
with a finite extra direction, in which {\it gauge invariance is broken
in a particular way}.  As a check, we calculate
the consistent anomaly in the three dimensional model and show that the
correct result is obtained.

\vspace{2ex}

  We begin in sect.~2 with a discussion of the axial anomaly in vector-like
gauge invariant models. We do so because the use of chiral defect
fermions for lattice simulations of QCD is an important subject by itself,
and because understanding the relation between the the vector-like
and chiral models may help in making further progress on the difficult
question of lattice chiral gauge theories.

  In a lattice formulation of QCD using chiral defect fermions
there is a natural way to define the axial current.
We calculate the axial anomaly using this definition.
We show that, up to a  numerical factor, this calculation can be reduced
to the calculation of the Chern-Simons action. Using the
result of Golterman, Jansen and Kaplan~[7], we find the coefficient of the
axial anomaly for any $n$.

  In sect.~3, after a discussion of the ambiguities of the infinite
lattice chiral model,
we introduce the {\it gauge variant} chiral model on a lattice with a finite
extra direction. We show that the carefully defined Feynman rules of the
infinite lattice model are reproduced by the gauge-variant model for a
sufficiently large extra dimension. We then calculate the consistent anomaly
in the gauge-variant model for the three dimensional case.
As explained below, although conceptually similar, the general case is
technically more complicated.

  In sect.~4 we discuss the behaviour of the gauge-variant
model in the presence of dynamical gauge fields. We argue that dynamical
restoration of gauge invariance takes place~[14], and that, very likely,
it is accompanied by the appearance of a new massless charged fermion
in the spectrum which makes the theory vector-like~[9,10].

\vspace{2ex}

  In most of this paper we will use the {\it boundary fermions}
variant~[8] of chiral defect fermions. Models with boundary fermions
are simpler, yet they have the same physical content as the corresponding
domain wall models. We will explain how the
same results can be obtained with domain wall fermions.

  Chiral fermions arise  on the $2n$ dimensional boundaries of
a $2n+1$ dimensional lattice model of Wilson fermions with a {\it constant}
mass $M$. They exist provided {\it free}
boundary conditions are chosen in the extra direction, and $M$ is in the
allowed range. (We set the Wilson parameter to $r=1$).
 Models with a single boundary (semi-infinite extra direction)
as well as models with two boundaries (finite extra direction)
can both be discussed.

  The chiral spectrum is the same as in the domain
wall case for the same choice of the absolute value of the mass.
Additional considerations~[4]  restrict the range of $M$ to $0<M<1$.
For this choice every boundary (like every domain
wall) supports a single chiral fermion.

  It is important to note that the relative sign of $M$ and $r$ in the
boundary fermion model is the opposite from their relative sign
in the conventional Wilson action. Thus, contrary to conventional
Wilsonian QCD, the sum of the $2n+1$ dimensional Wilson term and
the mass term is not a positive operator. This property has profound
impact on the dynamics of the model and is crucial for the existence of
stable massless modes.

  In Kaplan's definition of the domain wall model~[1],
the relative sign of $M$ and $r$ is the same as in the
Wilson action on the negative half-space. Again, the region
with unconventional relative sign -- the positive half-space -- is
in some sense more important. This can be seen from the fact that
the Chern-Simons current flows only into this half-space, while
it vanishes away from the wall on the other side. One can say
that the negative half-space can be discarded without affecting the
dynamics. This is precisely what one does in going from the domain
wall to the boundary fermions model.

\vspace{5ex}
\noindent {\bf 2.~~ Axial anomaly in the vector-like model}
\vspace{3ex}

  The vector-like model is defined as follows~[8]. We consider a $d=2n+1$
dimensional lattice. The first $2n$ coordinates, labeled $x_\m$, have an
infinite range, whereas the extra coordinate $s$ takes the values
$s=1,\ldots,2N$. We have chosen an even number of sites  to minimize
the amount of relabeling needed to see the correspondence between the
vector-like model and the chiral model discussed later.
None of our results depends on this choice. The action is
$$
  S=\sum_{x,y,s} \Bar\j(x,s) D^\parallel_{x,y}(U) \j(y,s) +
  \sum_{x,s,s'} \Bar\j(x,s) D^\perp_{s,s'} \j(x,s') \,,
\eqno(1)
$$
$$
  D^\parallel_{x,y}(U) = \half \sum_\m
  \left(
  (1+\g_\m) U_\m(x) \d_{x+\hat\m,y} +
  (1-\g_\m) U^\dagger_\m(x-\hat\m) \d_{x-\hat\m,y}
  \right) + (M-4) \d_{x,y}
\eqno(2)
$$
$$
  D^\perp_{s,s'} = \half (1+\g_d) \d_{s+1,s'} +
  \half (1-\g_d) \d_{s-1,s'} - \d_{s,s'}
\eqno(3)
$$
Notice that, apart from the unconventional sign of the mass term,
$D^\parallel_{x,y}(U)$ is the usual $2n$ dimensional gauge covariant Dirac
operator for Wilson fermions. The spectrum contains a right-handed fermion
near the boundary $s=1$ and a left-handed fermion near the other boundary.

  The action (1) is invariant under a $180^\circ$ rotation in the
$(k,d)$ plane around an axis located at the hyper-plane $s=N+1/2$.
Explicitly
$$
  \j(x_1,\ldots,x_{2n},s) \to i\g_k\g_d \,
  \j(x_1,\ldots,x_{k-1},-x_k,x_{k+1},\ldots,x_{2n},2N+1-s) \,.
\eqno(4)
$$

  For simplicity, we will consider below a U(1) gauge theory. The
generalization to the non-abelian case is straightforward. We now want to
define the various $2n$ dimensional currents. The vector current is uniquely
determined by the coupling to the gauge field, and it is
given by
$$
  J^V_\m(x)=\sum_{s=1}^{2N} j_\m(x,s) \,,
\eqno(5)
$$
where $j_\m(x,s)$ stands for the first $2n$ components of the $d$-dimensional
current
\beqrabc{6}
  j_\m(x,s) & = & \half\left(
        \Bar\j(x,s) (1+\g_\m) U_\m(x) \j(x+\hat\m,s)\right. -  \NON
    & & \left. \quad\quad
        \Bar\j(x+\hat\m,s) (1-\g_\m) U^\dagger_\m(x) \j(x,s)
                  \right)\,, \\
  j_d(x,s) & = &  \half\left(
                  \Bar\j(x,s) (1+\g_d) \j(x,s+1) -
                  \Bar\j(x,s+1) (1-\g_d) \j(x,s)
                  \right)\,.
\eeqr
The $d$-dimensional current satisfies the continuity equation
$$
  \sum_\m\D_\m\, j_\m(x,s) =
     \left\{ \begin{array}{ll}
          - j_d(x,1)\,,      &  s=1 \,,     \\
          - \D_d\, j_d(x,s)\,, &  1<s<2N-1 \,, \\
            j_d(x,2N-1)\,,      &  s=2N \,.
     \end{array}\right.
\eqno(7)
$$
Here $\D_\m f(x,s) = f(x,s)-f(x-\hat\m,s)$.
Notice the peculiar form of the boundary
terms in the continuity equation. Eqs.~(5) and~(7) imply the conservation
of the vector current.

  Next we want to define an axial transformation, from which
the axial current can be derived in the usual way. There is a lot of
arbitrariness in the choice of the axial transformation. Any transformation
that assigns opposite charges to the two chiral modes will reduce to the
usual axial transformation in the continuum limit.

  For example, we could define the axial transformation to be the usual
$2n$ dimensional one, applied equally to the fermions
on all $2n$ dimensional layers.
The disadvantage of this definition is that the divergence of
(singlet or non-singlet) axial
currents will involve an axial variation of the mass and Wilson terms,
summed over all $s$. It has been argued that QCD with chiral defect fermions
does not require any fine tuning of the mass term in the continuum limit~[8].
The verification of this statement is extremely complicated with the above
axial transformations, whereas it becomes trivial (within
the context of perturbation theory) with the axial transformation defined
below.

  The natural definition of an axial transformation in a model of chiral
defect fermions
takes advantage of the fact that the two chiral modes are globally separated
in the $s$-direction. We first define left-handed and right-handed
transformations as follows
\beqrabc{8}
  \d_{L,R} \j_{x,s}     & = & + i q_{L,R}(s) \j_{x,s} \,, \\
  \d_{L,R} \Bar\j_{x,s} & = & - i q_{L,R}(s) \Bar\j_{x,s} \,,
\eeqr
where
$$
  q_R(s) = \left\{ \begin{array}{ll}
              1\,,  &  1\le s \le N \,, \\
              0\,,  &  N <s \le 2N   \,,
           \end{array}\right.
\hspace{15mm}
  q_L(s) = \left\{ \begin{array}{ll}
              0\,,  &  1\le s \le N \,, \\
              1\,,  &  N <s \le 2N   \,.
           \end{array}\right.
\eqno(9)
$$

  Notice that the  two transformations are related by the discrete
symmetry of eq.~(4). The left-handed
and right-handed transformations act vectorially, but only on half of the
fermions, whereas fermions in the other half-space are invariant.
Thus, the right-handed massless mode at the $s=1$ boundary transforms only
under the right-handed transformation etc.
The non-invariance of the action under these transformations
resides entirely in the coupling between the $N$-th layer and the
$(N+1)$-st layer.

  We define the axial transformation to be the product of a right-handed
transformation and an inverse left-handed transformation.
Thus, fermions in the
two half-spaces transform with opposite charges under the axial
transformation. The corresponding currents are
\beqrabc{10}
  J^R_\m(x) & = & \sum_{s=1}^N j_\m(x,s) \,, \\
  J^L_\m(x) & = & \sum_{s=N+1}^{2N} j_\m(x,s) \,, \\
  J^A_\m(x) & = & J^R_\m(x)-J^L_\m(x) \,.
\eeqr
The divergences equations are
\beqrabc{11}
  \D_\m J^R_\m(x) & = & -j_d(N,x) \,, \\
  \D_\m J^L_\m(x) & = & j_d(N,x) \,, \\
  \D_\m J^A_\m(x) & = & -2j_d(N,x) \,.
\eeqr

   Feynman rules are obtained in the usual way by making the weak
coupling expansion
$$
  U_{x,\m}=\exp ig V_\m(x+\m/2) \,.
\eqno(12)
$$
Because of lack of translation invariance in the extra coordinate, we go
to momentum space only in the first $2n$ coordinates. The fermion propagator
$G_F(s,s';p)$ is given in ref.~[8]. (When necessary, the dependence of the
propagator on the size of the extra direction will be indicated by a
superscript).  For $s$ and $s'$ both near the same
boundary, $G_F(s,s';p)$ is dominated by the zero mode's contribution,
whereas for $s$ or $s'$  far from the boundaries,  $G_F(s,s';p)$ can be
approximated up to an exponentially small error by the translationally
invariant $d$-dimensional propagator
$$
  G_F(s,s';p_\m) \approx \int_{-\p}^\p {dp_d\over 2\p}\,
  e^{ip_d(s-s')} G_0(p_\a) \,,
\eqno(13)
$$
$$
  G_0(p_\a) =  { i\g_\a \hat{p}_\a + w(p_\a) - M \over
     \hat{p}_\a \hat{p}_\a + (w(p_\a)-M)^2 }    \,,
\eqno(14)
$$
where $p_\a=(p_\m,p_d)$ is the $d$-dimensional momentum (summation over
$\a$ is implied), $\hat{p}_\a = \sin p_\a$, and
$w(p_\a) = \sum_\a 1-\cos p_\a$.

  The single photon vertex is given by
$$
  \Tilde\L_\m(s,s';p,q,k) = i \d_{s,s'}\, \d^{2n}(k+p-q)\, \L_\m(p+q) \,,
\eqno(15)
$$
Here $p$ and $q$ are the $2n$ dimensional momenta of the incoming and
outgoing fermions respectively, and $k$ is the momentum of the external
gauge field.  $\L_\m$ is the usual $2n$ dimensional vertex
$$
  \L_\m(p+q)= \g_\m \cos \left( {p_\m+q_\m\over 2} \right)
    -i\sin \left( {p_\m+q_\m\over 2} \right)\,.
\eqno(16)
$$

  In addition, the Feynman rules include an integration over a $2n$
dimensional Brillouin zone for every closed loop, and a summation over
$s=1,\ldots,2N$ at each vertex.

  We are now ready to compute the divergence of the axial current in the
presence of an external gauge field. For a finite lattice spacing one expects
$$
  \D_\m J^A_\m = C\!_A\, \e_{\m_1\ldots \m_n,\n_1\ldots \n_n}
  (\partial_{\m_1}V_{\n_1})\cdots (\partial_{\m_n}V_{\n_n}) + O(a) \,.
\eqno(17)
$$
We want to calculate the coefficient $C_A$. Using eq.~(11c),
this is obtained from the
correlator of $j_d(N,x)$ with $n$ vector currents. Following ref.~[7]
closely, we Taylor expand this diagram. Keeping track of the various
symmetry factors we find
$$
  C\!_A={ i \e_{\m_1\ldots \m_n,\n_1\ldots \n_n} \over (2n)! }
      \left.
      {\partial\over\partial (p_1)_{\m_1}}\cdots
      {\partial\over\partial(p_n)_{\m_n}}
      T^A_{\n_1\ldots \n_n} (p_1,\ldots,p_n)
      \right|_{p_i=0} \,,
\eqno(18)
$$
$$
  T^A_{\n_1\ldots \n_n} = -2\int_{-\p}^\p {d^{2n}k \over (2\p)^{2n}}
  \sum_{s_1\ldots s_n}
  \left( \ct^+_{\n_1\ldots \n_n} - \ct^-_{\n_1\ldots \n_n} \right) \,,
\eqno(19)
$$
\beqn{20}
  \ct^{\pm}_{\n_1\ldots \n_n} & = &
  \pm \half \tr (1\pm\g_d)\,G_F(N+\h_\mp,s_1;k)\,
  \L_{\n_1}(k+k_1)\,G_F(s_1,s_2;k_1) \cdots \NON
  & & \quad\quad\cdots \L_{\n_n}(k_{n-1}+k_n)\, G_F(s_n,N+\h_\pm;k_n)\,.
\eeqn
Here $\h_+=1$, $\h_-=0$, and $(k_i)_\m=k_\m+(p_1)_\m+\ldots+(p_i)_\m$.

  Thanks to the exponential damping of the fermionic propagator in the
$s$-direction, only $s$-values in the neighbourhood of $s=N$ are
important on the \rhs of eq.~(19). For sufficiently large $N$ we can
therefore
replace $G_F$ in eq.~(20) by the translation invariant propagator $G_0$,
and extend the range of the $s$-summations to $-\infty<s<\infty$.
This gives rise to a considerable simplification, and allows us to express
$T^A_{\n_1\ldots \n_n}$ as an integral over a $2n+1$ dimensional Brillouin
zone. The resulting expression coincides with the corresponding one
in ref.~[7] up to a numerical factor. Using the result of ref.~[7] for the
coefficient of the Chern-Simons action we find
$$
  C\!_A = { 2i (-)^{n+1} \over (2\p)^n\, n! } \,,
\eqno(21)
$$
in agreement with known results~[15].

  If one works with domain wall fermions, the vector-like model is defined
by imposing periodic boundary conditions. One lets $s$ range from $-2N$ to
$2N$ with the layers $s=2N$ and $s=-2N$ identified, the domain wall is at
$s=0$ and the anti-domain wall at $s=2N$. The right-handed transformation is
applied to the half-space $-N < s \le N$, and the left-handed
transformation to the other half-space.
The divergence of right-handed current
is $ j_d(-N,x) -j_d(N,x)$, with similar expressions for the other
currents. As shown in ref.~[7], the new term makes no contribution to the
anomaly.

\vspace{5ex}
\noindent {\bf 3.~~ Consistent anomaly in the chiral model}
\vspace{3ex}

  We now procced to discuss the infinite lattice chiral model. We will
first describe the naive Feynman rules of this model and exhibit their
ambiguities. Next we will explain how a unique consistent prescription for
evaluating the infinite $s$-summation is obtained.
We then show that the infinite lattice chiral model is really  the
limit of a sequence of gauge-variant models with increasing extra
direction. Finally we calculate the consistent anomaly in the three
dimensional case and find the correct result.

  The boundary fermion version of the chiral model is formally defined
by replacing the finite range of $s$ by a semi-infinite one $s\ge 1$.
With this change in the range of $s$, the action is still given by
eqs.~(1-3). The two dimensional source current is
$$
  J_\m(x)=\sum_{s=1}^\infty j_\m(x,s) \,.
\eqno(22)
$$
The continuity equation for the $2n+1$ dimensional current $j_\a(x,s)$
takes the standard form for all
$s\ge 2$, while for $s=1$ it is  given by the first row of eq.~(7).
If one is not careful about the limiting procedure implicit on the \rhs
of eq.~(22), one find that the {\it formal} divergence of the source current
is zero. It is our purpose to examine this question more carefully below.

  The Feynman rules undergo the following modifications. In each vertex,
the $s$-summation now extends over all $s\ge 1$. Also, the fermion
propagator has a different form~[8], which exhibits the presence of a single
chiral fermion near the single boundary at $s=1$.

 In more detail, if we restrict our attention to any finite interval
$1\le s,s' \le N$  and take $N'\gg 1$,
than the semi-infinite lattice propagator $G_F^\infty$ is
well approximated by the propagator $G_F^{N+N'}$ of a lattice with a
finite extra direction $1\le s \le N+N'$.
In fact, there is a uniform bound
$$
  \left| G_F^\infty(s,s';p) - G_F^{N+N'}(s,s';p) \right| \le
  c_1 e^{-c_2 N'}/|p|\,,
  \quad\quad 1\le s,s' \le N\,,
\eqno(23)
$$
for some calculable
positive constants $c_1$ and $c_2$. In eq.~(23) we have explicitly
shown the momentum dependence of the bound. For exponentially small momenta
the bound is useless.  This momentum range can however be disregarded
because, for an appropriate definition of the continuum limit,
it can be made exponentially small also compared to any
{\it physical} scale of the interacting theory.

  On the other hand, the two propagators differ
radically for $p\ll 1$ (in lattice units) and $s,s'\leqx N+N'$.
In this range,
$G_F^{N+N'}$ propagates a massless state whereas $G_F^\infty$ does not.
Another obvious difference is that $G_F^{N+N'}$ is not even defined outside the
range $1\le s,s' \le N+N'$. This, however, is less important because
$G_F^\infty$ does not propagate any light states for large $s$.
Any spurious effect of the existence of infinitely may heavy
fields can be cancelled by the introduction of appropriate Pauli-Villars
fields~[4,11].

  The infinite $N$ limit is therefore singular. This singularity is
reflected in the ambiguities of the naive Feynman rules of the infinite
lattice model, to which we now turn.

  The example we will consider is the vacuum polarization in the three
dimensional model. The naive Feynman rules of the infinite lattice model
give rise to the following expression
$$
  \P_{\m\n} = \int_{-\p}^\p {d^2 k \over (2\p)^2}
  \sum_{s_1,s_2=1}^\infty I_{\m\n} \,, \quad\quad\quad \mbox{(formally)}
\eqno(24)
$$
$$
  I_{\m\n} = \L_\m(-2k-p)\,G_F^\infty(s_1,s_2;k)\,
             \L_\n(2k+p)\,G_F^\infty(s_2,s_1;k+p)\,.
\eqno(25)
$$

  Unfortunately, the \rhs of eq.~(24) is ambiguous. In order to exhibit
this ambiguity let us consider the IR regularized quantity
$$
  \P_{\m\n}^{N_1,N_2} = \int_{-\p}^\p {d^2 k \over (2\p)^2}
  \sum_{s_1=1}^{N_1} \sum_{s_2=1}^{N_2} I_{\m\n} \,.
\eqno(26)
$$
Notice that $\P_{\m\n}^{N_1,N_2}$ is the correlator
$$
  \P_{\m\n}^{N_1,N_2}=\svev{J_\m^{N_1}\, J_\n^{N_2} } \,,
\eqno(27)
$$
where the regularized current is
$$
  J_\m^N = \sum_{s=1}^N j_\m(x,s) \,.
\eqno(28)
$$
Here $N$ stands for $N_1$ or $N_2$.

  The regularized current satisfies a divergence equation analogous to
eq.~(11a). In momentum space we find
\beqrabc{29}
  \P_\n^{N_1,N_2} & \equiv & 2\sum_\m \sin (p_\m/2)\,
                                        \P_{\m\n}^{N_1,N_2} \\
   & = & -  \int_{-\p}^\p {d^2 k \over (2\p)^2}
           \sum_{s_2=1}^{N_2} \left(I_\n^+ - I_\n^- \right)\,,
\eeqn
$$
  I_\n^\pm =   \pm \half \tr (1\pm\s_3)\,G_F(N_1+\h_\mp,s_2;k)\,
  \L_\n(2k+p)\,G_F(s_2,N_1+\h_\pm;k+p) \,.
\eqno(30)
$$
Notice that eq.~(30) coincides with
eq.~(20) of the vector-like model for $n=1$.

  Let us now consider several limiting procedures and check the
transversality of the resulting expression. We define
\beqrabc{31}
  \P_\n   & = &
          \lim_{N_1\to\infty}\left.\P_\n^{N_1,N_2}\right|_{N_1=N_2}\,, \\
  \P'_\n  & = &
          \lim_{N_2\to\infty}\lim_{N_1\to\infty} \P_\n^{N_1,N_2}\,, \\
  \P''_\n & = &
          \lim_{N_1\to\infty}\lim_{N_2\to\infty} \P_\n^{N_1,N_2}\,.
\eeqr
Every one of these limiting procedures gives a different result.
Using the exponential damping of the propagator in the $s$-direction,
the reader
can easily verify that $\P'_\n=0$. $\P''_\n$ is recognized as the
divergence of the right-handed current in the vector-like theory (see
eq.~(19) and the following discussion), which is equal to one half of the
axial anomaly.

  The correct limiting procedure is the one in eq.~(31a). The limiting
procedures~(31b) and~(31c) are unacceptable because they involve a
regularized vacuum polarization which does not obey Bose symmetry. A
glance at eq.~(27) reveals that, except for $N_1=N_2$, $\P_{\m\n}^{N_1,N_2}$
describes the coupling of two photons to two {\it different} source
currents, which is obviously incorrect. Insisting that the  regularized
vacuum polarization describe the coupling of two photons to the {\it same}
source current implies the limiting procedure~(31a). More generally,
the correct prescription for evaluating any Feynman graph is the following.
One first performs all the $s$-summations over a finite range
$1\le s_1,s_2,\ldots \le N$. Only at the end the common largest value $N$
is sent to infinity.

  We will soon extract the two dimensional consistent anomaly from $\P_\n$.
But first we show that the Feynman
rules of the infinite lattice chiral model, as carefully defined above,
identify it with the limit of a sequence of {\it gauge-variant} lattice
models with a finite extra direction.

  The gauge-variant model consists of $N$ layers of charged fermions
and $N'$ layers of neutral fermions. Gauge invariance is broken in the
coupling between the last layer of charged fermions and the first layer
of neutral fermions. The action is (compare eq.~(1))
\beqn{32}
  S & = & \sum_{x,y}\sum_{s=1}^{N}
             \Bar\j(x,s) D^\parallel_{x,y}(U)\j(y,s) \NON
    & & + \sum_{x,y}\sum_{s=N+1}^{N+N'}
             \Bar\j(x,s) D^\parallel_{x,y}(1)\j(y,s) \NON
    & & + \sum_x \sum_{s,s'=1}^{N+N'}
             \Bar\j(x,s) D^\perp_{s,s'} \j(x,s') \,,
\eeqn

  The tree level spectrum consists of a charged right-handed fermion
at the $s=1$ boundary, and a neutral left-handed fermion at the
boundary $s=N+N'$. The source current is given by eq.~(28).
The $s$-summation in every vertex
extends only over the limited range $1\le  s_1,s_2,\ldots \le N$.
At this stage, the similarity to the correct Feynman rules of the infinite
lattice model is already clear. To make it precise, we make use of the
bound~(23). This bound is applicable because we need the propagator only
in the above limited range. As a result, in the limit $N'\to\infty$
one can replace the finite lattice propagator $G_F^{N+N'}$
by $G_F^\infty$, thus recovering the regularized form of the diagrams
of the infinite lattice chiral model.

  We comment that, unlike the ambiguous situation described earlier,
the precise order of the limits $N\to\infty$ and $N'\to\infty$ is
unimportant. The reason is that for any large value of $N'$ one
avoids the dangerous region where the finite and infinite lattice
propagators disagree, and this is true regardless of the value of $N$.
In particular, one can choose $N=N'$. For this choice, the
correlator of any number of source currents in the gauge-variant model
is equal to the correlator of the same number of right-handed currents in
the vector-like model provided the external gauge field is switched off.

  Finally, let us calculate the consistent anomaly of the three dimensional
model. Recall that at the linearized level, the consistent anomaly is
given by an expression similar to the \rhs of eq.~(17), but with the
constant $C\!_A$ replaced with another constant $C_{cons}$.
We thus have to evaluate the relevant part of $\P_\n$ using eqs.~(29),
(30) and~(31a) in the limit of large $N$. We try to apply the
same reasoning as in the calculation of the axial anomaly in the vector-like
model. In eq.~(29b), only the region $s_2\approx N$ contributes to
the $s_2$-summation (we let $N_1=N_2=N$),
and so we can replace the finite lattice
propagator by the translation invariant propagator $G_0$. For the same
reason, the lower limit of the $s_2$-summation can be extended to $-\infty$.
However, the upper limit lies in the center of the region $s_2\approx N$
and it cannot be altered. Making the substitution $s=s_2-N$ we thus obtain
$$
  \P_\n =  -  \int_{-\p}^\p {d^2 k \over (2\p)^2}
        \sum_{s=-\infty}^0 \left( \hat{I}_\n^+ - \hat{I}_\n^- \right) \,,
\eqno(33)
$$
$$
  \hat{I}_\n^\pm =   \pm \half \tr (1\pm\s_3)\,G_0(\h_\mp,s;k)\,
  \L_\n(2k+p)\,G_0(s,\h_\pm;k+p) \,.
\eqno(34)
$$

  Up to this point, everything could be trivially generalized to an arbitrary
dimension. Thus, we are able to express the consistent anomaly using
translationally invariant propagators, but summations that extend only
over a semi-infinite extra direction. One can go to momentum representation
also in the extra direction by making use of the Fourier transform of
the lattice $\th$-function. However, we have not found this representation
very useful in the computation of the diagram.

  In the three dimensional case we proceed as follows. By making a
$180^\circ$ rotation in the $(2,3)$ plane we find $\P_\n=\Tilde\P_\n$, where
$$
  \Tilde\P_\n = -  \int_{-\p}^\p {d^2 k \over (2\p)^2}
        \sum_{s=1}^\infty \left( \hat{I}_\n^+ - \hat{I}_\n^- \right) \,.
\eqno(35)
$$
Notice that the difference between $\P_\n$ and $\Tilde\P_\n$ is in the
range of the $s$-summation.
We comment that for the symmetric choice $N=N'$, this rotation is
essentially the discrete symmetry eq.~(4). However, the limiting
expressions eqs.~(33) and~(35) are related through this rotation also for
an asymmetric limit where $N\ne N'$.

  We complete the computation
by invoking the relation to the vector-like model. Specifically,
$\P_\n$ and $\Tilde\P_\n$ correspond to the correlators
$\svev{\D_\m J_\m^R \, J_\n^R}$ and $\svev{\D_\m J_\m^R \, J_\n^L}$ of the
vector-like model respectively. Using  $J_\m^V=J_\m^R+J_\m^L$,
the equality of the above two correlators
and eq.~(11), we find that in three dimensions $C_{cons}=C\!_A/4$.
Using eq.~(21) for $n=1$ we find $C_{cons}=i/4\p$ as expected.

  We comment that for $n>1$,
the above discrete symmetry is not sufficient to reduce the calculation of
the consistent anomaly to that of the axial anomaly, (which, as we have
seen, can be related to the calculation of the Chern-Simons action).
This is somewhat disappointing because the Chern-Simons action in $2n+1$
dimensions is closely related to the consistent anomaly in $2n$ dimensions.
However, we stress that the difficulties we have encountered are of a
technical nature, and we have every reason to believe that similar results
should exist for any $n$.

\vspace{5ex}
\noindent {\bf 4.~~ Discussion}
\vspace{3ex}

  A model of chiral defect fermions on a lattice with a
(semi)-infinite extra direction is formally both gauge invariant and
chiral. An arbitrary chiral spectrum should give rise to
anomalies. On the other hand, a manifestly gauge invariant lattice action
implies exact current conservation. The conflict between these two
properties raises the suspicion
that the infinite lattice chiral model is not well defined.
Indeed, we have explicitly demonstrated the ambiguities of its Feynman rules.

  In a carefully defined model, either gauge invariance or the chiral
character of the spectrum are lost. We have shown above that
the carefully defined (semi)-infinite lattice model is actually the limit of
gauge-variant models with a finite
extra direction. These models consist of a charged and a neutral $2n+1$
dimensional slabs, each with a finite extra direction. If the two slabs were
decoupled, there would be one massless mode of each chirality on every slab.
The direct coupling between the two slabs breaks gauge invariance and,
at the same time, eliminates one charged and one neutral field from the
massless spectrum.

  In an attempt to construct chiral gauge theories on the lattice,
giving up exact gauge invariance at the lattice scale is a reasonable price,
if one can show that the spectrum remains chiral and gauge invariance is
recovered in the continuum limit  for anomaly free theories.
There are strong indications, however, that this scenario is not
realized in the gauge-variant model we have constructed.
When the gauge fields are promoted to full-fledged dynamical
variables, what is likely to
happen is dynamical restoration of gauge invariance, accompanied
by the appearance of new massless charged fields which make the theory
vector-like.

  The action of an arbitrary lattice model with a dynamical gauge field can
be written as $S=S_{inv}+S_{non}$, where $S_{inv}$ and $S_{non}$ are the
gauge invariant and gauge variant parts of the action respectively. (We are
allowing for the possibility that one of these terms is zero). Using gauge
invariance of the lattice measure, the partition function can be rewritten
as a functional integral over the original fields plus a fixed radius Higgs
field $\f$ taking values in the gauge
group~[14]. The action becomes $S=S_{inv}+S_{non}(\f)$,
where every field in
$S_{non}(\f)$ undergoes the gauge transformation defined by the field $\f$.

  If the above procedure is applied to the action~(32), the quadratic
gauge variant terms turn into gauge invariant Yukawa interactions. A model
of this type has been recently investigated in ref.~[9]. If the Yukawa
coupling is switched off, the charged and neutral fermions decouple and the
model is vector-like. This model could have a chiral continuum limit if it
had a strongly interacting symmetric (PMS) phase, in which the charged and
neutral massless modes that couple through the Yukawa interaction become
a massive Dirac field, while the other charged field remains massless.
Although the results of ref.~[9] are not conclusive,
the cumulative evidence points to the absence of such a phase, and
hence that the vector-like spectrum persists throughout the entire phase
diagram.

  More generally, we have recently derived a No-Go theorem~[10] that
extends the Nielsen-Ninomiya theorem to interacting lattice theories with
arbitrary couplings. Apart from unitarity and gauge invariance, which are
common to both theorems, the key assumption of ref.~[10] is that the decay
rate of a certain two point function (the retarded anti-commutator)
at space-like separations satisfies a mild bound. When this
bound is satisfied, one can show that the inverse retarded propagator,
considered as an effective hamiltonian, is sufficiently differentiable to
apply the Nielsen-Ninomiya theorem.

  Lattice models lacking reflection positivity are in general not unitary
at the lattice scale, yet they can have a perfectly consistent continuum
limit. An examination of the details of our theorem reveals that this kind
of non-unitarity is irrelevant, because in the proof one uses only
unitarity of the low energy spectrum. In particular, the inclusion of heavy
Pauli-Villars fields in a model of chiral defect fermions should not affect
the validity of the No-Go theorem since they automatically decouple from
low energy physics.

  What makes the application of our No-Go theorem to a given lattice model
less straightforward than in the case of the Nielsen-Ninomiya theorem,
is the need to verify the validity of the above mentioned bound.
Various arguments indicate that many (and perhaps all)
short range lattice models should satisfy this bound. At the moment, however,
direct evidence is limited to the free field case and to strongly interacting
models amenable to certain analytical methods.

  In conclusion,
our purpose in the present paper was to show that the infinite lattice
model of chiral defect fermions, when properly defined, is in fact a
special case of the gauge invariant model proposed in ref.~[3].
Further study along the lines of refs.~[9] and~[10]
is needed before a definite conclusion can be drawn about the
feasibility of a chiral continuum limit in this model.

  Part of this research was carried out during a visit to DESY.
I thank I.~Montvay, M.~Luscher and the other members of the Theory Group at
DESY  for their hospitality and for stimulating discussions.

\vspace{5ex}
\centerline{\bf References}
\vspace{3ex}
\newcounter{00001}
\begin{list}
{[~\arabic{00001}~]}{\usecounter{00001}
\labelwidth=1cm}

\item D.B.~Kaplan, \PLB{288} (1992) 342.

\item K.~Jansen, \PLB{288} (1992) 348. K.~Jansen and M.~Schmaltz,
\PLB{296} (1992) 374.

\item D.B.~Kaplan, \NPBP{30} (1993) 597.

\item R.~Narayanan and H.~Neuberger, \PLB{302} (1993) 62.

\item C.P.~Korthals-Altes, S.~Nicolis and J.~Prades, {\it Chiral
defect fermions and the layered phase}, CPT--93/P.2920.
J.~Distler and S--J. Rey, preprint PUPT--1386.
S.~Aoki and H.~Hirose {\it Perturbative analysis for Kaplan's lattice
chiral fermions}.

\item R.~Narayanan and H.~Neuberger, preprints RU--93--25 and RU--93--34.

\item M.F.L.~Golterman, K.~Jansen and D.B.~Kaplan, \PLB{301} (1993) 219.

\item Y.~Shamir, preprint WIS--93/20--FEB--PH, to appear in \NPB.

\item M.F.L.~Golterman, K.~Jansen, D.N.~Petcher and J.C.~Vink,
preprint UCSD/PTH 93--28, Wash.~U. HEP/93--60.

\item Y.~Shamir, preprints WIS--93/56--JUNE--PH,
hep-lat/9306023 and WIS--93/57--JULY--PH, hep-lat/9307002.

\item S.A.~Frolov and A.A.~Slavnov, preprint MPI-Ph 93--12.

\item W.~Bock, J.~Smit and J.C.~Vink, preprint ITFA 93--18, UCSD/PTH 93--15.

\item H.B.~Nielsen and M.~Ninomiya, \NPB{185} (1981) 20,
{\it Erratum} \NPB{195} (1982) 541; \NPB{193} (1981) 173.

\item D.~Foerster, H.B.~Nielsen and M.~Ninomiya, \PLB{94} (1980) 135.
S.~Aoki, \PRL{60} (1988) 2109. K.~Funakubo and T.~Kashiwa, \PRL{60} (1988)
2113.

\item P.H.~Frampton and T.W.~Kephart, \PRD{28} (1983) 1010.

\end{list}

\end{document}